\documentclass[journal=jacsat,manuscript=article]{achemso}
\usepackage{bibentry}
\usepackage{chemformula} 
\usepackage[T1]{fontenc} 
\usepackage{soul} 
\usepackage{amsmath,bm}

\usepackage{natbib}

\usepackage{setspace}
\usepackage[version=3]{mhchem} 
\usepackage[T1]{fontenc}       
\usepackage{lineno}

\author{Hinaut Antoine}
\affiliation{Department of Physics, University of Basel, Klingelbergstrasse 82, 4056 Basel, Switzerland}
\email{antoine.hinaut@unibas.ch}
\author{B. Sena Tömekçe}
\affiliation{Physics Department E20, TUM School of Natural Sciences, Technical University of Munich, Garching, Germany}
\author{Shuyu Huang}
\affiliation{Department of Physics, University of Basel, Klingelbergstrasse 82, 4056 Basel, Switzerland}
\author{Yiming Song}
\affiliation{Department of Physics, University of Basel, Klingelbergstrasse 82, 4056 Basel, Switzerland}
\author{Ernst Meyer}
\affiliation{Department of Physics, University of Basel, Klingelbergstrasse 82, 4056 Basel, Switzerland}
\author{Antonio Cammarata}
\email{cammaant@fel.cvut.cz}
\affiliation{Department of Control Engineering, Faculty of Electrical Engineering, Czech Technical University in Prague, Technicka 2, 16627 Prague 6, Czech Republic}
\author{Willi Auwärter}
\email{wau@tum.de}
\affiliation{Physics Department E20, TUM School of Natural Sciences, Technical University of Munich, Garching, Germany}
\author{Thilo Glatzel}
\email{thilo.glatzel@unibas.ch}
\affiliation{Department of Physics, University of Basel, Klingelbergstrasse 82, 4056 Basel, Switzerland}

\title{Superlubricity of Borophene: Tribological Properties in Comparison to hBN}

\abbreviations{AFM, ncAFM, STM, UHV}

\begin{document}
\begin{abstract}
The tribological performance of 2D materials makes them good candidates toward a reduction of friction at the macroscale. Superlubricity has been observed for graphene, MoS\textsubscript{2} and MXenes and hexagonal boron nitride (hBN) is used to reduce or tune friction, but other materials are investigated as potential candidates for low-lubricity applications. Specifically, borophene is predicted to have ultra-low friction.
Here, we experimentally investigate frictional properties of borophene and use a borophene-hBN lateral heterostructure to directly compare the tribological properties of the two complementary 2D materials. In particular, we investigate the friction between a sliding tip and (i) the weakly corrugated $\mathcal{X}_6$-borophene layer on Ir(111) or (ii) the hBN/Ir(111) superlattice structures with a strongly corrugated moiré reconstruction. Our experimental study performed in ultra-high vacuum at room temperature combined with a Prandtl-Tomlinson (PT) model calculation confirms the superlubricity predicted for borophene, while hBN, which exhibits a higher friction, is nevertheless confirmed as a low friction material. Ab initio calculations show that the lower friction of $\mathcal{X}_6$-borophene with respect to hBN can be rationalized by weaker tip/surface interactions. In addition, we assess structural and electrical properties of borophene and hBN by using scanning probe techniques and compare their dissipation under the oscillating tip to investigate the possible path of energy dissipation occurring during friction. 
Our study demonstrates the low frictional properties of borophene and the potential of lateral heterostructure investigations to directly compare the properties of these 2D materials. 
\end{abstract}

\section{Introduction}

The tribological properties of 2D materials favor their use as solid lubricants \cite{berman15,berman18,wyatt21,marian22}. Although selected 2D materials are already incorporated into devices, fundamental studies, down to the atomic scale, are needed to understand the origin of their frictional properties  \cite{lee10,choi11,mandelli17,song20,song22,song22a,gnecco24}. In particular, the atomic structures, electronic properties, and interactions with the support can have a major influence on their tribological performance. For monolayers, possible moiré structures are known to strongly influence the lubricity \cite{liu23,song22}. Chemical modification, defects, or grain boundaries further influence lubricity \cite{kwon12,zambudio21,song24}. Furthermore, measurements of lateral heterostructures are of interest to directly compare the frictional behavior of 2D materials\cite{vazirisereshk19a,cai22}.

Among 2D materials, borophene, a monolayer of boron atoms, has gained particular attention due to its interesting properties, such as electrical conductivity, mechanical strength, and chemical reactivity. First synthesized on an Ag(111) surface\cite{mannix15,feng16}, borophenes by now have been prepared on various supports \cite{ou21,innis25,li25}. As theoretically predicted and experimentally confirmed, different polymorphs can be formed, depending on the growth method, the support material, or other treatments that can influence the desired properties \cite{kiraly19,liu19,vinogradov19,wang19,cuxart21,kaneti22,kumar24}. The potential of borophene as tribological materials has been anticipated early on \cite{sachdev15} and mechanical characteristics were modeled \cite{tsafack16,zhang17a,shukla18,yang25,sun25}. Nonetheless, respective experimental studies are largely missing. Specifically, outstanding tribological performances such as ultra-low friction and even a negative friction coefficient are so far only theoretically predicted \cite{xu21,xu22,xu22b,xu23a} and yet to be proven experimentally. 
Hexagonal boron nitride (hBN), a stoichiometric mixture of boron and nitrogen atoms arranged in a honeycomb lattice, is established as lubricant or as an intercalation layer, given its tribological performance \cite{kimura99,mandelli17,song18}.  The formation of moiré patterns for surface-supported monolayer hBN adds interesting electronic and mechanical properties~\cite{koch12} and allows hBN to be used as nano-templated support for atoms or molecules~\cite{auwarter19}. 

Here, we grow lateral heterostructures to directly investigate and compare the structural, electrical and tribological properties of borophene and hBN on Ir(111). The structures of the layers, e.g. $\mathcal{X}_6$-borophene/Ir(111) and moiré formation for hBN/Ir(111), are probed by scanning tunneling microscopy (STM) and non-contact atomic force microscopy (nc-AFM) at room temperature in ultra-high vacuum (UHV), complementing former low-temperature STM characterizations~\cite{cuxart21}. We access the work function of both layers with Kelvin probe force microscopy (KPFM) as well as their dissipation in the presence of an oscillating tip with nc-AFM.   
Their tribological properties are then investigated via friction measurements using contact AFM in comparison with calculations using the Prandtl-Tomlinson (PT) model. The ultra-low friction of the borophene layer is revealed experimentally for the first time. The hBN layer exhibits higher friction and remains wear-free.
The better tribological performance of the borophene layer is rationalized by an atomistic model: while the tip/surface interaction is predicted to be weaker for $\mathcal{X}_6$-borophene compared to hBN, at the same time the relatively higher friction of the hBN layer seems also to be related to the corrugated moiré pattern and its energy dissipation channel. Our results demonstrate the predicted interesting frictional properties of borophene and show the potential of lateral heterostructure measurements to directly compare 2D materials. 

\section{Results and discussion}

\subsection{The lateral borophene-hBN interface}

\subsubsection{Borophene and hBN structures}
\begin{figure*}[!ht]
	\centering
	\includegraphics[scale=1]{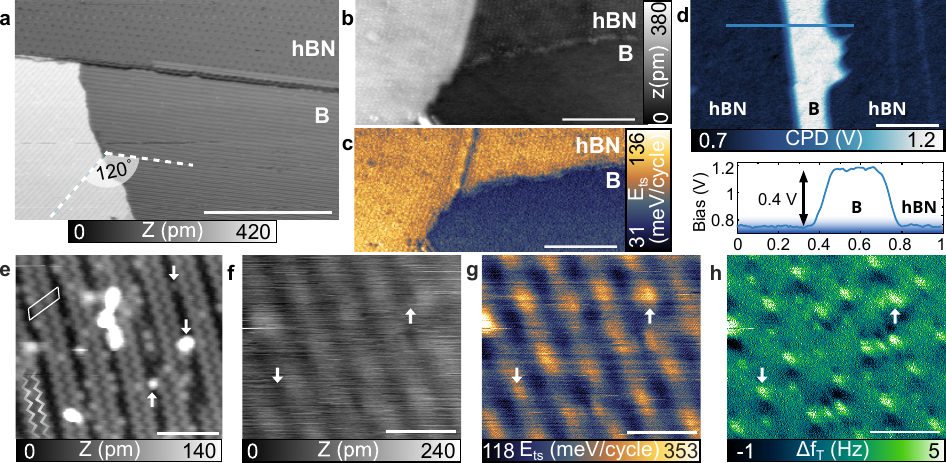}
	\caption{Large-scale borophene-hBN interface. a) STM topography. b) nc-AFM topography and corresponding c) dissipation in meV per oscillation cycle. d) CPD and corresponding profile. e) High resolution STM topography. f) High resolution nc-AFM topography and corresponding g) dissipation (meV per oscillation cycle) and h) torsional frequency shift. Parameters:  a) $I = 200$~pA, $U = -0.3$~V. b,c) $f_0 = 164$~kHz, $A = 4$~nm $\Delta f = -17$~Hz. e) $I = 90$~pA, $U = 1$~V. f,g,h) $f_0 = 169$~kHz, $A = 2$~nm $\Delta f = -130$~Hz, $f_t = 1.53$~MHz, $A_t = 80$~pm. Scale bars a,b,c) $50$~nm. d) $500$~nm. e,f,g,h) $3$~nm.}
	\label{fig1}
\end{figure*}

Lateral heterostructures of borophene (denoted by B in all figures) and hBN are grown by dosing diborane and borazine onto a preheated ($\simeq 1200$~K) clean and atomically flat Ir(111) surface maintained under UHV conditions, following a previously reported method\cite{cuxart21} (see Methods section for details). 
Borophene and hBN islands are obtained on the Ir(111) surface, with a total coverage close to a monolayer, as visible in the STM topography image of Fig.~\ref{fig1}a. Borophene and hBN domains with a size of more than hundred nanometers are formed. A hexagonal moiré structure is observed for the hBN domains\cite{koch12,farwickzumhagen16,schulz18,auwarter19} (see Fig.~S2 for high resolution). 
The borophene islands show a striped pattern characteristic for the $\mathcal{X}_6$ polymorph on Ir(111), in agreement with studies at cryogenic temperatures\cite{vinogradov19,cuxart21,omambac21,kamal23}. On a large scale this can be seen as rows indicated by the dotted lines in Fig.~\ref{fig1}a, showing two, of the three possible, rotational domains (120$^\circ$ between row orientation) on the surface. The borophene domains do not grow over Ir(111) step edges, as can be seen from the change in orientation when crossing the monoatomic Ir(111) step. When formed on the same terrace, borophene and hBN have an apparent  height difference of $~35$ pm at -0.3 V, in line with previous observations \cite{cuxart21} (see Fig.~S1a for height profiles).

The large-scale lateral heterostructure of borophene-hBN on Ir(111) is also imaged by nc-AFM measurements, Figs.~\ref{fig1}b and c show the topography and the simultaneously measured dissipation. The hBN islands are located on two Ir(111) terraces, where the lower terrace is shared by hBN and borophene. The hexagonal moiré structure of hBN is visible in both topography and dissipation. Height profiles showing the monoatomic step ($\approx 210$~pm) of Ir(111) below hBN and the flat interface between borophene and hBN (here no height difference measured) are provided in Fig.~S1. The hBN layers separated by the Ir(111) step-edge exhibit identical dissipation, as visible in the dissipation image of Fig.~\ref{fig1}c. The borophene layer shows a much lower dissipation, in meV per oscillation cycle, than the hBN. Such lower dissipation is a good indicator of the out-of-plane rigidity and adhesion of the borophene on the Ir(111) surface in comparison to the hBN on Ir(111). X-ray photoelectron spectroscopy (XPS) measurements have shown that borophene is strongly interacting with Ir\cite{kamal23}, with an interaction exceeding the one of hBN with Ir(111) \cite{vinogradov19,farwickzumhagen16}. The hBN layers are known to have a strong tendency to deform under an oscillating tip\cite{koch12}. 

\subsubsection{Borophene electronic properties}

We use KPFM in the frequency modulation mode to measure the contact potential difference (CPD) of the $\mathcal{X}_6$-borophene and hBN layers on Ir(111). Both borophene and hBN islands can be clearly distinguished in the CPD images, as shown in Fig.~\ref{fig1}d. The borophene island shows a higher CPD, corresponding to a higher work function than the hBN layer with a difference of $400$~meV \cite{sadewasser18}.  Using the bare Ir(111) work function as a reference, we can estimate absolute work function values for hBN and borophene, see Fig.~S2 for more information. Assuming a work function of $\simeq 5.78$~eV for Ir(111)\cite{derry15,liu19}, we evaluate the work functions of $\mathcal{X}_6$-borophene and hBN to be 4.68~eV and 4.28~eV, respectively. The value for borophene is close to the work function reported for a freestanding borophene sheet (4.75~eV) and different borophene polymorphs on Ag(111)\cite{liu21a}, but smaller than the one recently reported for borophene on Ir(111) with a potential influence of adsorbates (5.30~eV)\cite{kamal23}. The work function of hBN is also in good agreement with previous field emission resonance experiments ($4.2-4.6$~eV)\cite{schulz14}. The work function difference between borophene and hBN might have an effect on the electron dissipation channel and could influence the overall friction behavior\cite{liu15,wolloch18,liu22}.  

\subsubsection{High resolution STM and AFM measurements}
The borophene $\mathcal{X}_6$ reconstruction is more clearly seen in high resolution STM images, as shown in Fig.~\ref{fig1}e. The stripped pattern is observed and the characteristic "wavy" appearance of the rows is visible (see white overlay zigzag shape). The measured $\mathcal{X}_6$ unit cell dimension of $1.65$~nm $\times$ $0.60$~nm with an internal angle of $60^{\circ}$ matches literature values\cite{vinogradov19,cuxart21,omambac21}, see Fig.~S3 for dimension and profiles. Structural defects are observed with the appearance or disappearance of some rows as well as more local defects (white arrows) \cite{cuxart21}.

The topography measured by nc-AFM on a $\mathcal{X}_6$-borophene island also reveals the presence of the rows, Fig.~\ref{fig1}f. Whereas in STM the "wavy" substructure of the rows is resolved, this is not the case in the nc-AFM measurements. Here, only the row structure is visible, with a measured width of $1.65$~nm, see Fig.~S3 for a profile. The simultaneously measured dissipation (Fig.~\ref{fig1}g) and the torsional frequency shift ($\Delta f_T$, Fig.~\ref{fig1}h) images also show the row structure of the borophene island. In the $\Delta f_T$ signal, the white protrusions observed between the rows (white arrows) are attributed to the defects and adsorbates  that are also visible in the STM topography~\cite{wang17,huang20}. 
In terms of dissipation, the lowest values are obtained on the areas identified as rows in the nc-AFM topography. The higher dissipation in between the rows could be induced by the presence of the defects. A spatial analysis of the structures in the different images is provided in Fig. S2.

The wavy pattern of the $\mathcal{X}_6$-borophene lattice visible in our STM measurements, is mainly due to the electronic interaction at the borophene/Ir(111) interface\cite{vinogradov19,omambac21}. In nc-AFM, a technique more sensitive to the topography of the surface, less details of the electronic structures are observed, also when using torsional imaging\cite{liu21,song22a}. The lower dissipation of the borophene compared to hBN suggests a flatter and more rigid borophene layer compared to the more deformable moiré structure formed by hBN\cite{koch12,koch13}. This is also shown by calculating the dissipation ratio ($Dr$) between the highest and lowest dissipation over borophene and hBN layers respectively. We found $Dr_{B} \simeq 3$ and $Dr_{hBN} \simeq 5.4$, indicating larger variation over hBN than borophene, in good agreement with their deformability. 

\subsection{Friction properties of borophene and hBN}

\subsubsection{Ultra-low friction of $\mathcal{X}_6$-borophene}
\begin{figure*}[!ht]
	\centering
	\includegraphics[scale=1]{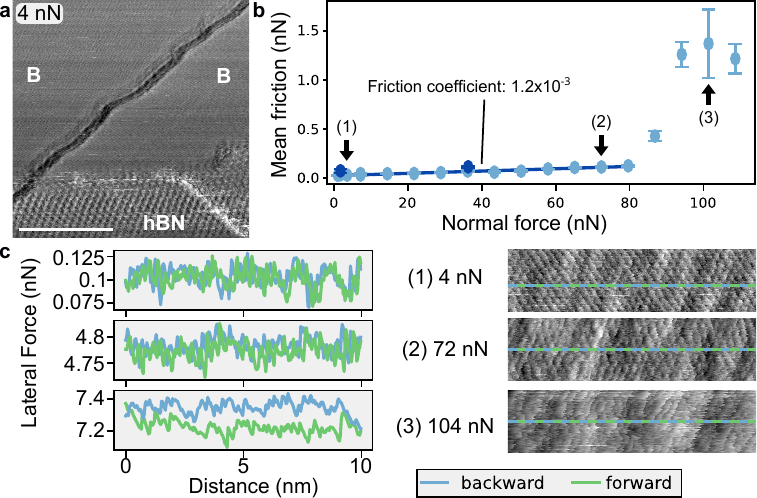}
	\caption{Borophene friction properties. a) Lateral force map of the borophene hBN layers on Ir(111) at an applied load of 4~nN.  b) Mean friction force of the borophene layer. c) Lateral force traces (left) and corresponding images (right) for applied loads of 4, 72 and 104~nN respectively. $\nu$ = $49$~nm/s. Scale bar a) 40~nm, image are $10$~nm wide in c).} 
	\label{fig2}
\end{figure*}

$\mathcal{X}_6$-Borophene and hBN domains are also distinguished by contact measurements, as shown in the forward lateral force image in Fig.~\ref{fig2}a. The $\mathcal{X}_6$-borophene is extended over two terraces in the upper part, while the hBN is detected in the lower part, as apparent from the moiré pattern. By changing the normal force, we performed load dependent frictional force measurements on these borophene~($\mathcal{X}_6$) islands as shown in Fig.~\ref{fig2}b. The mean frictional forces are calculated by recording the forward and backward traces following method described in ref\cite{meyer21}.   The $\mathcal{X}_6$-borophene exhibits superlubricity\cite{berman18}, as indicated by the calculated coefficient of friction of $1.2 \times 10^{-3}$ for the lower loads. Such superlubric sliding is maintained up to a load value of $80$~nN. For higher loads, the friction force becomes non-linear and superlubricity disappears. However, by reducing the normal force back below the threshold, the superlubricity can be reversibly restored. This is exemplified by the two measurements taken after the high loads (dark blue points in Fig.~\ref{fig2}b). The recovery of the low mean friction force value is an indication that both the tip and the surface are preserved during the friction experiments (up to the maximum applied load).  
The lateral force images and traces show the transition from the superlubric to the dissipative friction regime as visible in Fig.~\ref{fig2}c, extracted from the mean friction measurement in Fig.~\ref{fig2}b (indicated by arrows). There, for low load and load just before transition (labels (1) and (2) on the graph), the forward and backward traces have no hysteresis. Atomic stick-slip is observed on borophene in the traces and atomic feature resolution is obtained in the lateral force images. At higher loads, without superlubricity (3 in the graph), a hysteresis is visible in the trace and the atomic features are not observed in the lateral force image, indicating a frictional regime. The ultra-low friction maintained over a wide range of normal loads, and it's recovery after higher loads, shows the excellent frictional properties of the borophene layer. 

\subsubsection{Friction of the borophene-hBN interface}
\begin{figure*}[!ht]
	\centering
	\includegraphics[scale=1]{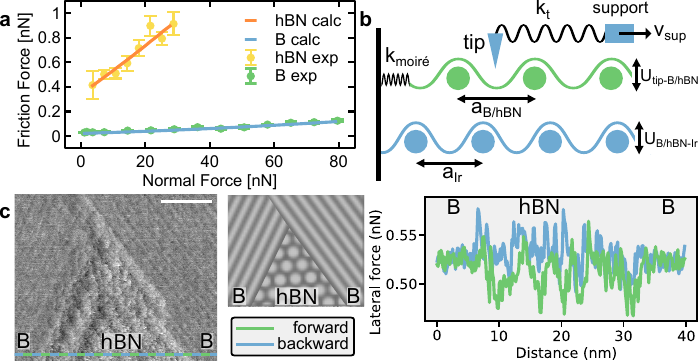}
	\caption{Friction properties of the (borophene-hBN)/Ir(111) lateral heterostructure. a) Friction force versus load for borophene and hBN. Dots are from experiments, lines from calculations. b) Scheme of the PT model used for the calculations. c) Lateral force image of the lateral borophene-hBN interface with the corresponding scheme and lateral force traces. Parameters: c). Scale bar $10$~nm  } 
	\label{fig3}
\end{figure*}

A direct comparison of the frictional properties of both the borophene and hBN islands allows us to confirm the very low friction of borophene. To this end, we obtained the mean friction value on the hBN layer with a similar load dependency measurement as shown in Fig.~\ref{fig3}a with the same tip. A friction coefficient of $2.0\times10^{-2}$ is obtained, showing the much higher friction on hBN compared to borophene. 
The difference between hBN and $\mathcal{X}_6$-borophene in terms of their frictional behavior can also be seen by direct comparison on a contact trace, sliding over both hBN and $\mathcal{X}_6$-borophene under constant load. This is visible in the contact image and the corresponding scheme in Fig.~\ref{fig3}c where an hBN island, embedded between two $\mathcal{X}_6$-borophene domains with different orientation, is scanned. The lateral force trace reveals the difference between the hBN and the borophene. On $\mathcal{X}_6$-borophene islands, i.e. on the sides, forward and backward scans yield similar values and no hysteresis is observed, implying reduced friction (compare values in Fig.\ref{fig2}). Above hBN, i.e. in the middle region, there is a hysteresis between forward and backward scans, indicating increased friction. The advantage of using a lateral heterostructure is clearly demonstrated here, with a direct comparison, in a single trace, revealing the higher friction of hBN compared to $\mathcal{X}_6$-borophene. 

\subsubsection{Prandtl-Tomlinson model applied at the interface}
To understand the difference in the lubricity regime between the two different 2D materials, we used a modified PT model~\cite{zhang22,huang25} on both borophene and hBN, including the effects of the moiré superstructure and its lateral flexibility. A scheme of the model and the parameters used is shown in Fig.~\ref{fig3}b. A tip is dragged at constant velocity by a spring over a sinusoidal potential representing the interaction between the tip and the borophene or hBN layers. The underlying Ir(111), with a different lattice constant, naturally gives rise to a moiré superlattice. The lateral deformability of the moiré interface during sliding is characterized by a spring, $k_{moire}$. The friction is determined by the product of tip spring constant $k_t$ and its tension between the tip apex position $x_{tip}$ and the support position $\nu_st$.  The potentials, with amplitude $U_{s}$ and periodicity $a_{s}$ are used to model the interaction between the tip and the 2D layers, where the subscript 's' (denoting substrate) should be specifically substituted with the corresponding material-specific parameters when describing borophene and hBN. We used $a_{B}$~=~$1.63$~\AA, $a_{hBN}$~=~$2.49$~\AA\ and $a_{Ir}$~=~2.71~\AA\  as lattice dimensions extracted from our experiments and adjusted the values of $U_{B}$ and $U_{hBN}$, which represent the corrugation of the tip-surface interactions, to fit the calculations to the experimental values. Further details of the calculation can be found in the supplementary information part~5.
To fit our calculations to the experimental values, we used a lower value for $U_{B}$ than for $U_{hBN}$. This reduced surface corrugation for borophene compared to hBN corresponds to a weaker energy barrier at the same normal force and it is in good agreement with experimental values, i.e. $U_{hBN}$ > $U_{B}$ indicates a more frictional surface.

\subsubsection{Potential energy surface of borophene and hBN}
\begin{figure*}[!ht]
	\centering
    \includegraphics[scale=1]{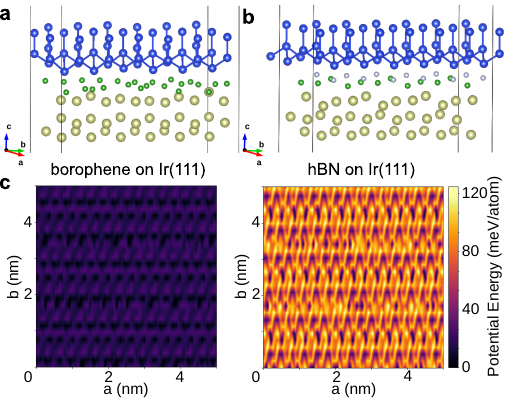}
	\caption{Schematic representation of the a) Ir-borophene-tip and b) Ir-hBN-tip geometric models used in the ab initio calculations. The golden, green, white and blue sphere represent the position of the Ir, B, N and Si atoms, respectively. c) Potential energy surface (PES) thermal map of borophene on Ir(111)  (left) and hBN on Ir(111) (right). }
	\label{fig:borhbn}
\end{figure*}

The experimental observation as well as the predictions of the PT model and the optimization with  $U_{hBN}$ >  $U_{B}$ are supported by our ab initio calculations. The strength of the tip-surface interaction has been evaluated by sampling the potential energy surface (PES) as a function of the position of the tip with respect to the surface. The PES scan has been realized by considering the model geometries for the $\mathcal{X}_6$-borophene and hBN system as in Fig.~\ref{fig:borhbn}. A $10\times 10$ grid sampling of the PES has been obtained by shifting the tip parallel to the $(\bm{a},\bm{b})$ plane with respect to the borophene or the hBN surface. We find that the potential energy barrier for the tip sliding over the surface is of $0.03$~eV/atom and $0.12$~eV/atom for the borophene and hBN systems, respectively (see Fig. \ref{fig:borhbn}c). This points out that it is easier to slide over borophene than over hBN.

We also evaluated the vertical displacements of the B and N atoms in the two systems. We find that, during sliding, the B atoms in the borophene layer displace in average by $0.15$~\AA{}, while the atoms in the hBN layer displace by 0.43~\AA{}. These results show that, during sliding, the borophene surface retains its flatness and has a weak interaction with the tip, while the hBN surface is more prone to deformations and has a strong interaction with the tip. Such results are in agreement with our STM-AFM measurements as discussed above.

\section{Conclusion}
We have experimentally demonstrated the superlubricity of borophene for the first time. We took advantage of the possibility to create lateral heterostructures of borophene and hBN to directly compare their mechanical, electronic and tribological properties. Using STM and nc-AFM at room temperature in UHV, we first characterized the borophene and hBN layers on the Ir(111) surface. We then measured a lower friction over borophene compared to hBN, which was further confirmed by both PT model and ab-initio calculations, showing a lower energy barrier for borophene, i.e. a lower friction force. The ab-initio calculation also revealed the much higher vertical displacement of the hBN layer compared to borophene, which can be compared to the much higher friction measured experimentally. Our results demonstrate the potential of lateral heterostructure measurements to directly compare 2D material properties.

\subsection{Materials and methods}

\textbf{Borophene and borophene-hBN heterostructure growth.}
The Ir(111) single crystals were prepared by repeated cycles of sputtering (Ar$^+$ ions at an energy of $1$~keV) and annealing at $1250$~K. Borophene and borophene-hBN heterostructures were grown on Ir(111) by dosing diborane and/or borazine while keeping the substrates at high temperature\cite{cuxart21}. To promote the growth of the borophene-hBN lateral heterostructures we dosed a diborane-borazine mixture, controlled by mass spectrometry, on a sample kept at $1200$~K. To promote borophene, we dosed a diborane-borazine mixture onto the sample at $1350$~K, resulting in the formation of pure borophene\cite{omambac23}. The annealing is maintained for 10~min after the end of the dosing. 

\textbf{Scanning probe experiments.}
STM, nc-AFM and contact AFM measurements were performed with a home-built microscope operated at room temperature and controlled with a Nanonis RC5 electronics. PPP-NCL cantilevers (Nanosensors) were used as sensors for nc-AFM (typical resonance frequencies of $f_{1st}$~=~$160$~kHz and $f_{t}$~=~$1.5$~MHz, oscillation amplitude $2$ to $5$~nm and $80$~pm, respectively). PPP-CONT cantilevers (Nanosensors) were used for friction measurements. Cantilevers preparation consisted of an annealing for $1$~h at $400$~K followed by an Ar$^{+}$ sputtering for $2$~min at $1$~keV at an Ar$^{+}$ pressure of $3 \times 10^{-6}$~mbar. STM tips were made from Pt/Ir wire. The UHV system was maintained at a base pressure of $5 \times 10^{-11}$~mbar during the measurements. 

\textbf{PT model calculation.}
The dynamics of the friction system is described by the equation of motion: 
\begin{equation}
\begin{split}
    m_t \dot{x}_t+m_t {\mu}_t (\dot{x}_t-v_s) &=-\frac{\partial{V(x_t,t)}}{\partial{x_t}}+\zeta_t(t); \\
    m_{s} \dot{x}_s+m_{s} {\mu}_{s} \dot{x}_{s} &=-\frac{\partial{V(x_{s},t)}}{\partial{x_{s}}}+\zeta_{s}(t).
    \label{eq1}
\end{split}
\end{equation}
where $m_t$ and $m_{s}$ represent the effective mass of the tip and the locally deformed borophene or hBN. The damping coefficients of the corresponding motion ${\mu}_t$ and ${\mu}_{s}$ are set to a critical damping. $\zeta$ stands for a thermal noise term, describing the random effects of fluctuation satisfying the fluctuation-dissipation relation. The potential energy of the system $V$ includes four contributions: the interaction potential between tip and borophene/hBN, $V_{t-s}$ ;the substrate-Ir(111) interface potential, $V_{s-Ir}$; the cantilever spring potential and the elastic strain energy associated with the moir\'e deformation. This can be expressed as: 

\begin{equation}
    V=V_{t-s}+V_{s-Ir}+\frac{1}{2}k_t (x_t-v_st)^2+\frac{1}{2}k_{moire} x_{s}^2,
    \label{eq2}
\end{equation}
where the interaction potentials are explicitly defined as:
\begin{equation}
    V_{t-s}=U_{t-s}cos(\frac{2\pi(x_t-x_s)}{a_s}),
    \label{eq3}
\end{equation}
\begin{equation}
    V_{s-Ir}=U_{s-Ir}cos(2\pi(\frac{x_t}{a_{Ir}}-\frac{x_t}{a_s}+\frac{x_s}{a_s}).
    \label{eq4}
\end{equation}

The equations of motion are numerically solved using the fourth-order Runge–Kutta algorithm. To investigate load-dependent friction behavior, the amplitude of the interfacial potential amplitude is varied to modulates the normal load. The average friction under the corresponding corrugated potential is obtained by averaging the instantaneous friction over 82 cycles.

\textbf{Ab initio calculations.}
The starting point to build the computational models for the $\mathcal{X}$6-borophene and hBN systems are the atomic geometries reported in \cite{vinogradov19,farwickzumhagen16}, respectively.
In our settings, the Ir surface, the layer and the Si tip are arranged in the ($\bm{a},\bm{b})$ plane, while a 25~\AA{} vacuum slab has been added along the $\bm{c}$ direction orthogonal to the layer plane, in order to prevent interactions between periodically repeated images (Fig.~\ref{fig:borhbn}).
We performed Density Functional Theory calculations \cite{hohenberg64} as implemented in the \textsc{vasp} software \cite{kresse96,kresse96a}. To describe the atomic interactions we choose the Perdew-Burke-Ernzerhof (PBE) \cite{perdew97} and the vdW-DF2 \cite{lee10} energy functionals for the borophene and hBN systems, respectively.
The plane wave energy cutoff is set to 500 eV and the irreducible Brillouin zone is sampled with a $3\times1\times1$ Monkhorst-Pack mesh \cite{monkhorst76}. Self-consistent field and geometric relaxation loops are considered converged within a tolerance of 10$^{-8}$~eV and 0.01~eV/\AA{}, respectively.
The atom positions and the lattice parameters $\bm{a}$ and $\bm{b}$ of the as-built Ir/$\mathcal{X}$6-borophene/Si and Ir/hBN/Si models have been optimised and used as starting point for sampling the potential energy surface.
The PES scan has been realised by displacing the position of the Si atoms along the $\bm{a}$ and $\bm{b}$ axes with respect to the borophene and hBN layers, and relaxing the position of all the atoms forming the Ir/borophene/Si and Ir/hBN/Si interfaces; regarding the Si atoms, only their position along the $\bm{c}$ axis has been optimised, in order to prevent full relaxation which would restore the tip position to the starting point.

\begin{acknowledgement}
We thank Hermann Sachdev and Marc G. Cuxart for fruitful discussions.  E.M., A.H., S.H. and T.G. acknowledge funding from the Swiss National Science Foundation (SNSF, Nanocontrol 200021L-219983, 200020-188445, CRSII5-213533, and 200021-231373), the Werner Siemens Foundation (WSS) and the Swiss Nanoscience Institute (SNI). Y.S. acknowledges the support of the SNF(CRSK-2\_228934). A.C. acknowledges the support of the Czech Science Foundation (project No. 24-12643L), co-funding by the European Union under the project "Robotics and advanced industrial production" (reg. no. CZ.02.01.01/00/22\_008/0004590),  the Ministry of Education, Youth and Sports of the Czech Republic through the e-INFRA CZ (ID:90254), and the access to the computational infrastructure of the OP VVV funded project CZ.02.1.01/0.0/0.0/16\_019/0000765 “Research Center for Informatics”.
\end{acknowledgement}

\subsection{Authors contribution}
A.H. W.A. and T.G. planned the experiments. A.H. performed the SPM experiments. S.H. assisted for friction experiments and analysis. A.H and B.S.T performed CVD growth. S.H. performed the PT model calculation. A.C performed the ab initio calculations. All the authors discussed the data and the manuscript. 

\begin{suppinfo}


\begin{itemize}
  \item Additional figures and captions. 

\end{itemize}

\end{suppinfo}

\bibliography{BoroBiB}

\end{document}


\begin{abstract}

\end{abstract}

\subsection{Part 1: borophene-hBN lateral interface with STM and nc-AFM}
\begin{figure*}[!ht]
	\centering
	\includegraphics[width=115mm]{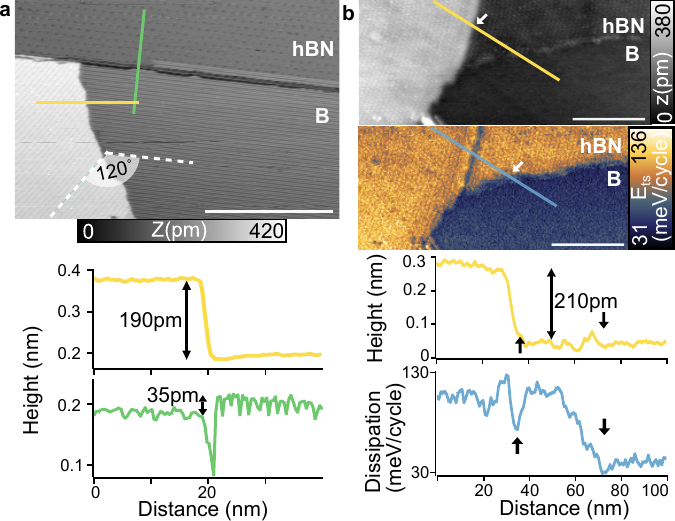}
	\caption{Borophene-hBN lateral interface on Ir(111) a) STM topography image and corresponding profiles. b) nc-AFM and corresponding dissipation images with profiles across lateral heterostructure. Parameters: a) $I = 200$~pA, $U = -0.3$~V. b) $f_0 = 164$~kHz, $A = 4$~nm $\Delta f = -17$~Hz.  Scale bars a,b) $50$~nm}
	\label{SIFig_B-hBNprofiles}
\end{figure*}
Using the STM topography over the lateral heterostructure Fig.~\ref{SIFig_B-hBNprofiles}a, the height of the Ir(111) step edge is measured to be 190 pm (below borophene). In nc-AFM a height of  210~pm (below hBN) is measured as visible in the topography of Fig.~\ref{SIFig_hBN}b. The difference in height between $\mathcal{X}_6$-borophene and hBN is measured to be 35 pm in STM and is flat with nc-AFM (Fig.~\ref{SIFig_hBN}a,b. 
The dissipation channel of nc-AFM experiment reveals a lower dissipation on the borophene islands than on the hBN (Fig.~\ref{SIFig_hBN}b). 

\newpage
\subsection{Part 2: hBN on Ir(111)}

\begin{figure*}[!ht]
	\centering
	\includegraphics[width=135mm]{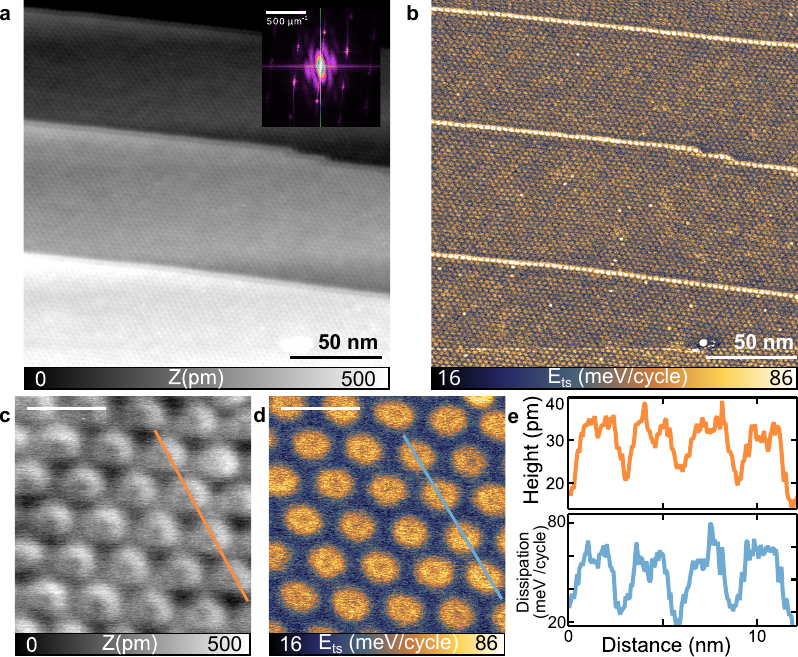}
	\caption{High resolution on hBN with nc-AFM. a) Large scale nc-AFM topography. Corresponding FFT in inset. b) Corresponding dissipation image in meV per oscillation cycle. c) Nc-AFM topography. b) Corresponding dissipation image in meV per oscillation cycle. e) Profiles from c) and d). Parameters: $f_0 = 169$~kHz, $A = 2$~nm, a,b) $\Delta f = -120$~Hz, c,d) $\Delta f = -130$~Hz. Scale bars a,b) $50$~nm, inset $500$~$\mu$m$^{-1}$, c,d) $5$~nm.} 
    \label{SIFig_hBN}
\end{figure*}

When dosing pure borazine on Ir(111) on a sample maintained at $1200$~K, a hBN monolayer is formed. Nc-AFM topography (Fig.~\ref{SIFig_hBN}a) and dissipation (Fig.~\ref{SIFig_hBN}b ) images over the hBN reveal the extension of the layer overs hundreds of nanometers as well as the moiré pattern. Fig.~\ref{SIFig_hBN}c,d are better resolved topography and dissipation images. A lattice close to $2.9$~nm is measured for the hBN moiré as reported in literature \cite{farwickzumhagen16,auwarter19}. 
Moiré structure and lattice dimension of the hBN islands in the mixed sample as shown in Fig.~1 are identical to here. 

\newpage
\subsection{Part 3: Borophene on Ir(111)}

\begin{figure*}[!ht]
	\centering
	\includegraphics[scale=1]{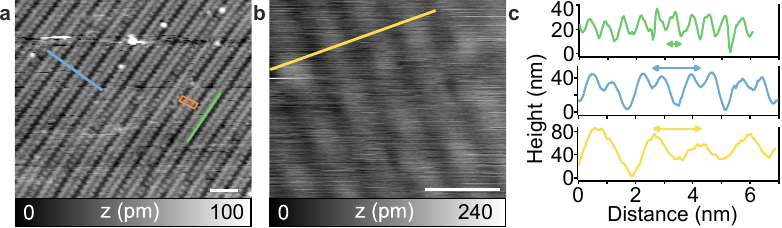}
	\caption{High resolution on $\mathcal{X}_6$ borophene-hBN. a) STM topography. b) nc-AFM topography. c) Profiles from a) and b). Parameters: a) $I=250$~pA, $U=-1$~V; b) $f_0 = 169$~kHz, $A = 2$~nm, $\Delta f = -120$~Hz, Scale bars a,b) $3$~nm.} 
     \label{SIFig_HR-borophene}
\end{figure*}

The dimensions of the $\mathcal{X}_6$-borophene are confirmed with profiles acquired on the STM and nc-AFM topography images as seen in Fig.~\ref{SIFig_HR-borophene}. From STM (Fig.~\ref{SIFig_HR-borophene}a), both directions of the $\mathcal{X}_6$ are obtained via profiles along and across the row structure while with nc-AFM (Fig.~\ref{SIFig_HR-borophene}b), only the profile across the rows is possible. Corresponding profiles are visible in Fig.~\ref{SIFig_HR-borophene}c. The $\mathcal{X}_6$ lattice unit cell measured dimension is $1.65$~nm $\times$ $0.60$~nm with an internal angle of $60^{\circ}$ corresponding to the values reported in literature \cite{vinogradov19,cuxart21,omambac21}

\newpage
\subsection{Part 4: Work function of Borophene and hBN} 

\begin{figure*}[!ht]
	\centering
	\includegraphics[scale=1]{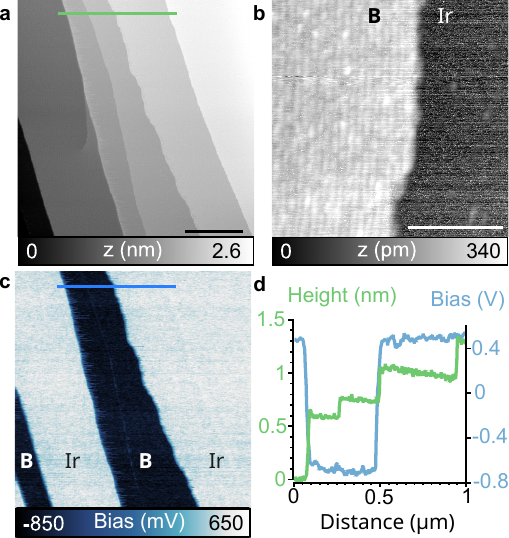}
	\caption{Borophene on Ir(111). a) NcAFM topography image of borophene grown on Ir(111). b) zoom topography image to reveal the $\mathcal{X}_6$-borophene reconstruction. c) CPD image simultaneous to a). Parameters: a-c) $f_0 = 169$~kHz, $A = 2$~nm $\Delta f = -130$~Hz. } 
	\label{SIFig_WF-borophene}
\end{figure*}

STM topography image over the $\mathcal{X}_6$-Borophene on Ir(111) surface. Showing both $\mathcal{X}_6$-borophene and Ir(111) area is shown in Fig.~\ref{SIFig_WF-borophene}a. A zoomed STM topography revealing the $\mathcal{X}_6$-Borophene rows is visible in Fig.~\ref{SIFig_WF-borophene}b. Using the Kelvin probe force image simultaneous to the large scale topography (Fig.~\ref{SIFig_WF-borophene}a) and the corresponding profiles, as seen in Fig.~\ref{SIFig_WF-borophene}c,d, we measure a contact potential difference of $1.1$~V between $\mathcal{X}_6$-Borophene and Ir(111). Using the known work function for the Ir(111) surface ($\simeq 5.78$~eV \cite{derry15,liu19}), the value of the $\mathcal{X}_6$-Borophene work function on Ir(111) is calculated to be WF$_{(\mathcal{X}_6-Borophene)/Ir(111)} = 4.68$~eV. 

Nc-AFM image corresponding to the KPFM image from Fig.~1d of the main manuscript. 
\begin{figure*}[!ht]
	\centering
	\includegraphics[scale=1.2]{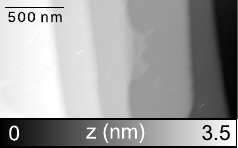}
	\caption{Borophene-hBN on Ir(111). Nc-AFM topography image of borophene hBN interface on on Ir(111) corresponding to Fig.~1 d) of manuscript.} 
	\label{SIFig_topo-borophene}
\end{figure*}

\newpage
\subsection{Part 5: Rows and defects in borophene with nc-AFM}

\begin{figure*}[!ht]
	\centering
	\includegraphics[width=120mm]{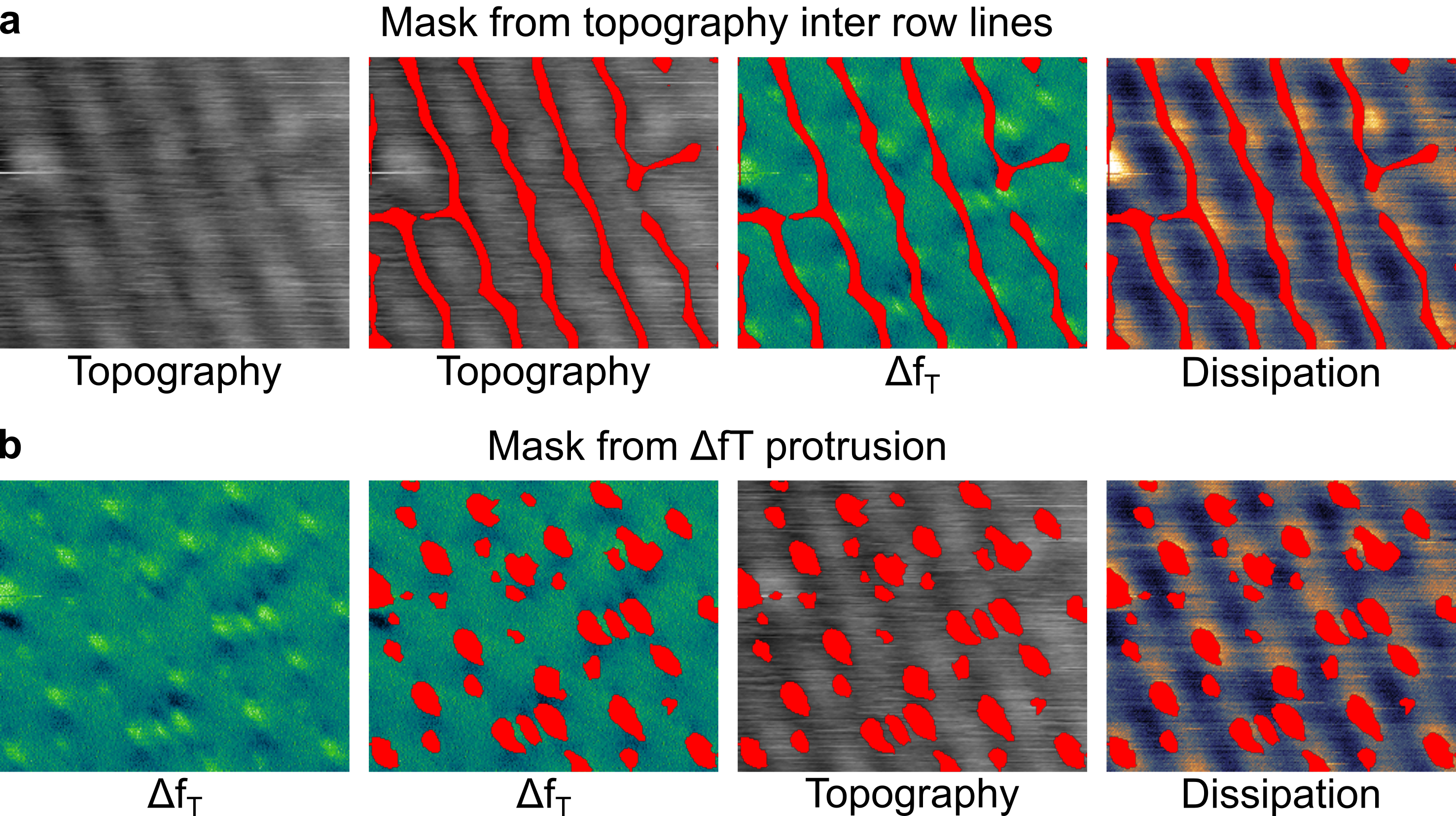}
	\caption{Rows and defects in borophene. a) Overlapping the mask from the inter rows lines from topography to $\Delta f_T$ and dissipation. b) Overlapping the mask from the defects in $\Delta f_T$ to topography and dissipation. $f_0 = 169$~kHz, $A = 2$~nm $\Delta f = -130$~Hz, $f_T = 1.53$~MHz, $A_T = 80$~pm. Image lateral size is $9$~nm} 
	\label{SIFig_overlap}
\end{figure*}

The rows structures and the defects in between them, as observed in nc-AFM topography and $\Delta f_T$ images, are superimposed to other signals to reveal their influence in the various signals in the Fig.~\ref{SIFig_overlap} and to allow correct identification of the rows structure of borophene in ncAFM. First, a mask (red) is created from the inter row lines observed in topography and then superimposed over the others signals (Fig.~\ref{SIFig_overlap}a). The same is done for the defects observed in $\Delta f_T$ (Fig.~\ref{SIFig_overlap}b).  

The defects observed in the $\Delta f_T $ signal are found in between the rows observed in the topography. This is similar to STM images, see Fig.~1e of the manuscript. Therefore the rows in nc-AFM coincide to the rows in STM. 

\newpage
\subsection{Part 6: The relationship between surface corrugation and normal force}

\begin{figure*}[!ht]
	\centering
	\includegraphics[width=135mm]{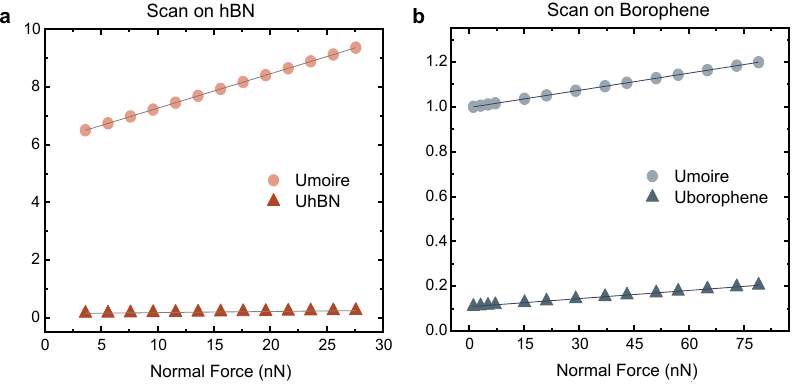}
	\caption{Relationship between surface corrugation and normal force. a) hBN. b) Borophene. }
	\label{SIFig_Ucalc}
\end{figure*}

\begin{table}
    \centering
    \begin{tabular}{lcr}
    1 & Torsional frequency of cantilever: $1551.6$~kHz & from exp.\\
    2 & Lateral stiffness of cantilever: $163.12937$~N/m & from exp. \\
    3 & Stiffness of moiré: $163.12937$~N/m & \\
    4 & Calculated mass of tip: $1.716 \times 10^{-10}$~kg & from exp. \\
    5 & Mass of locally deformed borophene/hBN: $1.716\times 10^{-9}$~kg & \\
    5 & a\textsubscript{hBN} = $2.49$~\AA & from exp \\
    6 & a\textsubscript{Ir} = $2.71$~\AA & from exp \\
    7 & a\textsubscript{B} = $1.63$~\AA & from exp\\
    8 & Damping coefficient of corresponding motions: & \\
     & $\eta_{1}$ = $2 \cdot \sqrt{k_{1}/m_{1}} \cdot 1 $ and $\eta_{2}$ = $2 \cdot \sqrt{k_{2}/m_{2}} \cdot 1 $& \\
    9 & Amplitude of corrugation potential: U\textsubscript{1} and U\textsubscript{2} & fit from exp \\
    10 & Temperature: $0$~K & \\
    \end{tabular}
    \caption{Parameters used for the calculation in the PT model and extraction information}
    \label{tab:my_label}
\end{table}


\bibliography{BoroBiB}